\documentstyle[12pt,epsfig]{article}
\def\reference{\parskip 0pt\par\noindent\hangindent 0.5 truecm}
\def\kms{km~${\rm s}^{-1}$}
\def\HI{H~{\sc i}} 
\def\HII{H~{\sc ii}}
\def\arcdeg{\hbox{$^\circ$}}
\def\arcmin{\hbox{$^\prime$}}

\def\fdg{\mbox{$.\!\!^\circ$}}%
\begin{document}
%
%
\title{\HI\ Emission and Absorption in the Southern Galactic Plane
Survey\footnote {Given as an oral presentation at ASA2K, Hobart,
Tasmania, July 2000}}
%


\author{N. M. McClure-Griffiths $^{1}$ \and
 John M. Dickey $^{1}$ \and
 B. M. Gaensler $^{2}$ \and
 A. J. Green $^{3}$ \and
 R. F. Haynes $^{4}$ \and
 M. H. Wieringa $^{4}$
} 

\date{}
\maketitle

{\center
$^1$ Department of Astronomy, University of Minnesota,  116 Church
St. S.E., Minneapolis, MN 55455, USA \\naomi@astro.umn.edu, john@astro.umn.edu\\[3mm]
$^2$ Center for Space Research, Massachusetts Institute of Technology,
70 Vassar St., Cambridge, MA 02139, USA \\bmg@space.mit.edu\\[3mm]
$^3$ Astrophysics Department, School of Physics, Sydney University,
NSW 2006 \\agreen@physics.usyd.edu.au\\[3mm]
$^4$ Australia Telescope National Facility, CSIRO, P.O. Box 76,
  Epping, NSW 2121 \\Raymond.Haynes@atnf.csiro.au, Mark.Wieringa@atnf.csiro.au\\[3mm]
 }

%
\begin{abstract}
We present preliminary results from the Southern Galactic Plane Survey
(SGPS) Test Region and Parkes data.  As part of the pilot project for
the Southern Galactic Plane Survey, observations of a Test Region
($325\fdg5 \leq l \leq 333\fdg5$; $-0\fdg5 \leq b \leq 3\fdg5$) were
completed in December 1998.  Single dish observations of the full
survey region ($253\arcdeg \leq l \leq 358\arcdeg$; $|b| \leq
1\arcdeg$) with the Parkes Radio Telescope were completed in March
2000.  We present a sample of SGPS \HI\ data with particular attention
to the smallest and largest scale structures seen in absorption and
emission, respectively.  On the large scale, we detect many prominent
\HI\ shells.  On the small scale, we note extremely compact, cold
clouds seen in \HI\ self-absorption.  We explore how these two classes
of objects probe opposite ends of the \HI\ spatial power spectrum.

\end{abstract}


\bigskip

%
%
\section{Introduction}
The Southern Galactic Plane Survey (SGPS) is an \HI\ spectral line and
$\lambda$21-cm continuum survey of the fourth quadrant of the plane of the
Milky Way (Dickey et~al.\ 1999).  The survey covers the region $253\arcdeg
\leq l \leq 358\arcdeg$; $|b| \leq 1\arcdeg$. The dataset, which is a
combination of data from the Australia Telescope Compact Array (ATCA) and
the Parkes Radiotelescope, is sensitive to angular scales larger than $\sim
1-2\arcmin$.  The ATCA data is a mosaic of 2212 pointings.  The Parkes data,
which was obtained with the inner seven beams of the multibeam receiver, has
extended coverage to $|b| \leq 10\arcdeg$.  The final data product includes
\HI\ data cubes with a spectral resolution of $0.8$ \kms\ and continuum maps
in all four Stokes parameters.

Observations for the SGPS are expected to be finished in late 2000 with a
full data release in early 2001.  Currently the Parkes observations are
complete, as are observations of the SGPS Test Region ($325\fdg5 \leq l \leq
333\fdg5$; $-0\fdg5 \leq b \leq 3\fdg5$).  This paper is based on a talk
given at the ASA Annual Meeting in July 2000.  Here we present a
representative sample of the SGPS \HI\ data, including low resolution data
from Parkes and full resolution data for the Test Region.  The continuum
emission has been subtracted from these \HI\ data.  A full description of
the observations and data analysis are discussed in detail elsewhere
(McClure-Griffiths~et~al.\ 2000a, 2000b).

In this introduction to the data we explore how the SGPS can contribute to
our understanding of the interstellar medium (ISM) over a large range of
size scales.  The SGPS data span several orders of magnitude in angular
size, allowing us to probe opposite ends of the \HI\ spatial power spectrum
over a significant portion of the inner Galaxy.  On the large scale we are
studying \HI\ emission structures, particularly shells and supershells to
understand the effect of these shells on the structure and dynamics of the
ISM.  The SGPS data may help us to understand how these deterministic
structures are ``pumping'' the spatial power spectrum.  On the small scale,
we are using \HI\ self-absorption (HISA) to probe the coldest, smallest
clouds in the ISM.  We seek to understand the ISM on all scales in between,
but the clues may be at the furthest extremes.

\section{\HI\ Shells}
We here present a small gallery of \HI\ shells detected in the Test
Region and the Parkes data.  These shells represent a selection of the
broad range of structures displayed by shells.  They range in diameter
from about 30 pc to 600 pc.  \HI\ shells provide an interesting
environment in which to study the ISM because they are among the
largest structures visible in the Galaxy.  In addition, they are among
the few discrete objects which are apparently deterministic in nature.
They result from relatively well understood phenomena such as a
supernova explosion for the smallest shells, or the combined effects
of stellar winds and many supernovae for the largest shells.  As the
fossils of extremely energetic events they allow us to study how
energy is injected into the ISM, and subsequently how that energy
affects the ISM as it undergoes the transition from deterministic to
turbulent in nature.

\subsection{Local Shell}
One of the smallest shells visible in the Test Region is extremely
local.  Figure~\ref{fig:local} shows a greyscale of a large angular
diameter shell centered at about $l=330\fdg5$, $b=2\fdg2$,
$v_{\sc{lsr}} \approx 2.1$ \kms.  Because of its local velocity,
distance estimates are extremely uncertain.  However, we used a
standard rotation curve for the Galaxy (Fich, Blitz \& Stark 1989), to
estimate the kinematic distance to this shell to be in the range 350
pc to 1 kpc.  At these distances the diameter is between 15 and 50 pc.
There is also tentative evidence for slow expansion on the order of 4
\kms, though no front or back caps are detected.  It should be noted,
though, that random \HI\ cloud motions are on the order of 6 \kms,
making it difficult to place much confidence in the expansion velocity
estimate.  The small size of this shell suggests that it may be an old
supernova remnant (SNR) which no longer radiates in the continuum.  It
is interesting to note, however, that the shell is extremely circular,
far more so than a typical SNR.
\subsection{Terminal Velocity Shell}
There is another small shell at the terminal velocity of $v=-110$ \kms.
This shell, as shown in Figure~\ref{fig:term}, is centered on $l=329\fdg9$,
$b=0\fdg4$ and is at a kinematic distance of $\sim 7.4$ kpc.  This distance
translates into a physical diameter of $\sim 51$ pc.  The shell stands out
as the only feature in the \HI\ channel maps near the terminal velocity that
persists over many velocity channels.  It is detectable from beyond the
terminal velocity at $v=-120$~\kms, where there is very little gas, to
$v=-105$~\kms, where it appears as a void surrounded by a great deal of \HI\ 
emission.  The structure remains changes very little over the entire
velocity range.  There are no obvious front or back caps and there is no
evidence of continued expansion.  With no obvious expansion it is difficult
to hypothesize about the origins of the shell.  Its size indicates that it,
too may be an old supernova remnant, but its curious position at the
terminal velocity suggests that it could be gas displaced from circular
rotation by any number of possible influences.
\subsection{RCW~94 Shell}
\begin{figure}[hbt] 
\begin{center}
\psfig{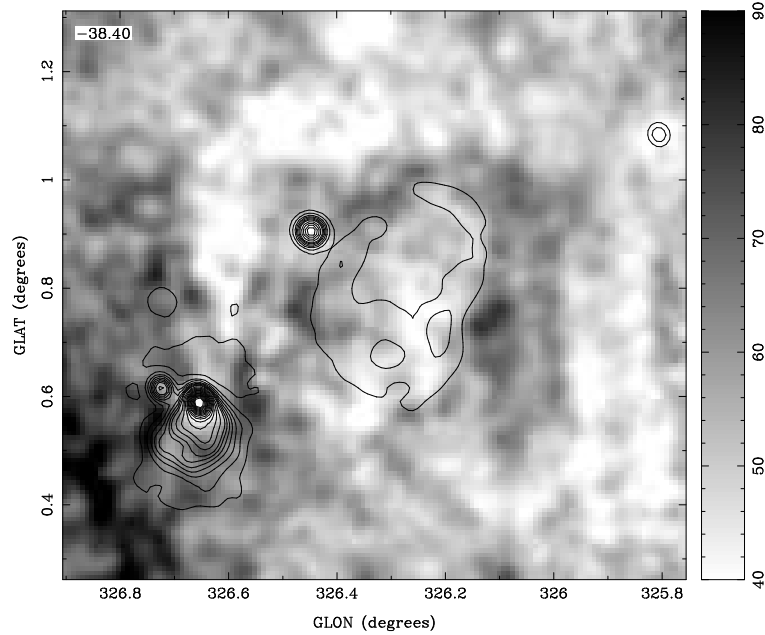}
\caption{Grey-scale plot of an \HI\ channel map at $v=-38$ \kms\
  with 21-cm continuum contours overlaid.  The grey scale is linear and runs
  from 0 to 80 K, while the contours are at 10 K intervals from 20 K to 400
  K.  The small shell, centered at $l=326\fdg3$, $b=0\fdg8$ is identified by
  the ring of emission around the \HII\ region.  The grey-scale is linear
  and is shown in the wedge on the right.}
\label{fig:rcw94}            
\end{center}
\end{figure}

There is an \HI\ shell surrounding the \HII\ region RCW~94 at
$l=326\fdg3$, $b=0\fdg8$.  The shell is itself surrounded by a ring
of \HI\ depletion.  Figure~\ref{fig:rcw94} shows the \HI\ channel map
at $v=-38$ \kms\ with continuum contours overlaid.  The \HI\ absorption
distance for this \HII\ region is $\sim 3.0$ kpc, which agrees well
with a distance of $\sim 3.1$ kpc determined by Caswell \& Haynes
(1987) using recombination line velocities.  The distance implies a
shell diameter of $\sim 25$ pc.  There is evidence for a small
expansion velocity of $\sim 5$ \kms.  We speculate that this shell was
formed in a molecular cloud, where the molecular hydrogen was
dissociated by the \HII\ region, RCW~94.  The depletion ring exterior
to the shell is most likely an effect of the high molecular gas
densities.
\subsection{Supershells}
Another class of \HI\ shells are so-called ``supershells'', with
diameters on the order of hundreds of parsecs.  Two large supershells
have been discovered in the Parkes data (McClure-Griffiths et al.\
2000b).  These shells, GSH 277+0+36 and GSH 280+0+59, have diameters of
620 pc and 430 pc, respectively, in addition to showing evidence of
break-outs above and below the Galactic plane (see
Figures~\ref{fig:gsh277} and \ref{fig:gsh280}).  These shells appear
to lie on the far edge of the Carina arm in the interarm region.  The
energy required to form a shell of this size is extremely large, on
the order of $10^{52}-10^{53}$ ergs.  At that level of energy input,
it is clear that supershells are one of the driving sources of energy
in the ISM and have a significant impact on the structure of the
Galaxy.
\section {\HI\ Self-Absorption (HISA)}
\begin{figure}[htb] 
\begin{center}
\psfig{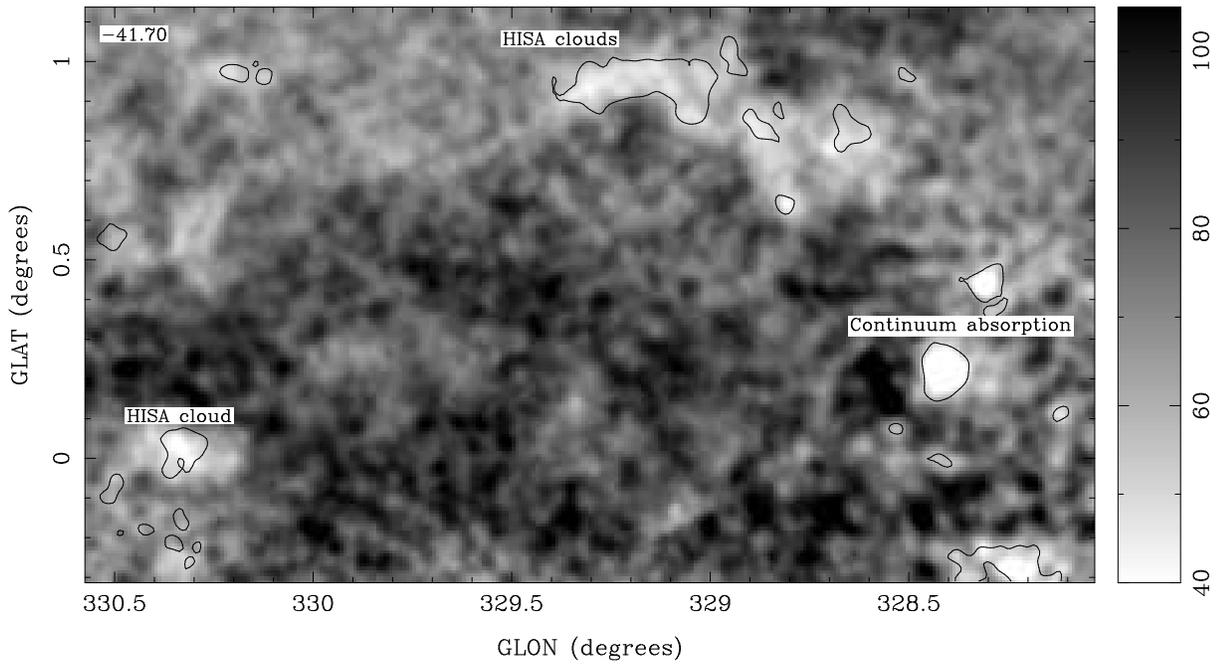}
\caption{This grey-scale plot is an \HI\ channel map at $v=-41.7$
\kms\ with a 55 K contour of the same channel overlaid.  The grey
scale is linear and runs from 40 to 105 K, as shown in the wedge on
the right.  The contours outline several prominent sources of
absorption, including a supernova remnant.  The extended areas marked
as HISA are cold clouds seen in absorption. }
\label{fig:hisa}            
\end{center}
\end{figure}

On the other end of the \HI\ spatial power spectrum we are using \HI\
self-absorption (HISA) to study the cold neutral medium (CNM).  HISA
occurs when cold foreground gas absorbs \HI\ emission from warmer
background gas at the same velocity.  The correspondence of one
velocity to two distances interior to the solar circle provides a
prime opportunity for studies of HISA.  HISA is characterized by
extremely narrow velocity width ($\sim 1-2$ \kms) absorption features
seen against the warmer background hydrogen emission.  The small, cold
clouds are also seen in the \HI\ channel maps as in
Figure~\ref{fig:hisa}.  New high-resolution \HI\ studies, such as the
Canadian Galactic Plane Survey (Taylor et al.\ 1999), are revealing
large numbers of cold clouds through HISA (Gibson et al.\ 2000).
Because HISA does not require background continuum sources, it is an
excellent way to study the cold neutral medium (CNM) throughout the
Galaxy.

\begin{figure}[htb] 
\begin{center}
\psfig{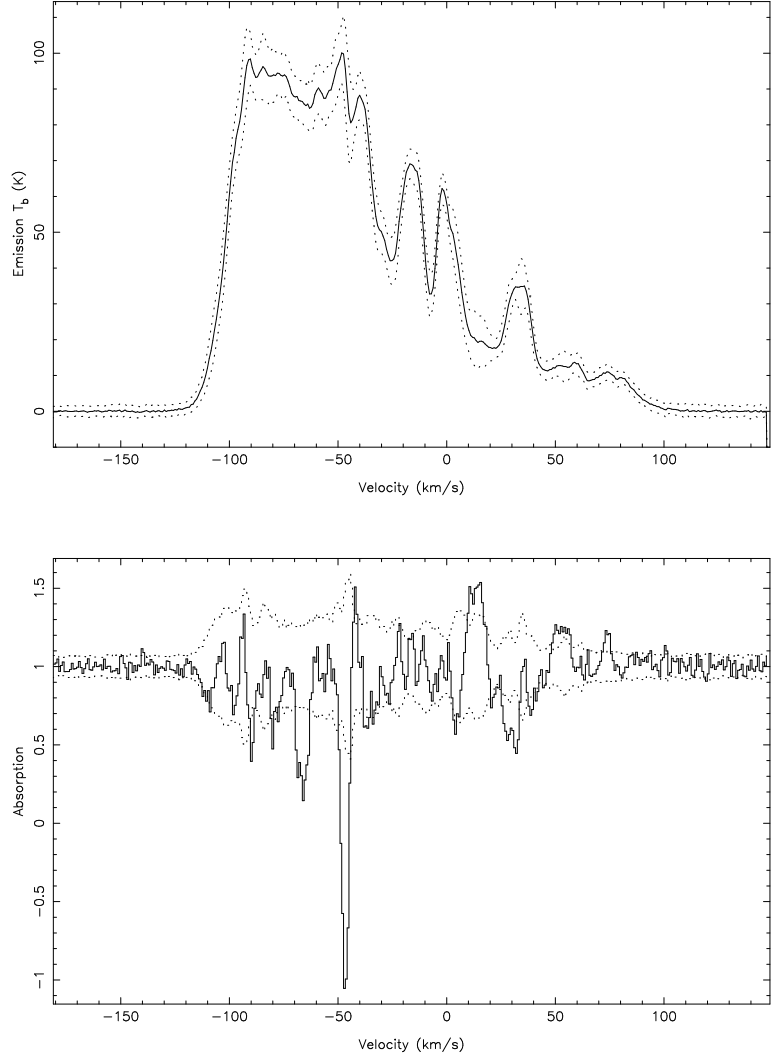}
\caption{The top panel is the emission profile near $l=330\arcdeg$
$b=0\arcdeg$.  The bottom panel shows the absorption profile for the
cold cloud seen at $l=330\fdg35$, $b=0\arcdeg$.  The dotted lines
trace the $1\sigma$ error envelope.  The narrow width of the
absorption feature is one of the defining characteristics of HISA.}
\label{fig:selfabs}            
\end{center}
\end{figure}

There are numerous HISA features in the SGPS Test Region.  These cold
clouds can be seen in the channel map shown in Figure~\ref{fig:hisa}.
An absorption profile towards the cloud at $l=330\fdg35$,
$b=0\arcdeg$ shows a characteristically narrow profile with a spin
temperature of $\sim 30$ K (Figure~\ref{fig:selfabs}).  Most of these
HISA features are in the velocity range $-50$ \kms\ $\leq v \leq -40$
\kms which, at these longitudes, corresponds to the edge of the
Scutum-Crux spiral arm.  We suggest that these cold clouds may mark
the compressed gas at the edge of the spiral shock, where molecular
clouds are likely to form.  While some HISA does clearly trace
molecular emission, other HISA are isolated, suggesting cold gas that
has not yet been compressed enough to form molecules (Gibson et al.\
2000).

The smallest and most dramatic features tend to stand out in initial
searches for HISA.  In order to study the HISA in a statistically
significant way, it will be necessary to develop non-subjective
searching techniques.  Such techniques have been developed for the
CGPS data by Gibson et al.\ (2000). Similar techniques are being
developed for the SGPS data.  The many cold clouds detected in the
CGPS and SGPS suggest excess emission fluctuations on the small side
of the spatial power spectrum.  It is possible that there are many
warmer, less dense clouds that we are not detecting.

\section{Questions and Future Work}
There are several questions to be answered about the \HI\ shells and
\HI\ self-absorption as we proceed with the SGPS.  With the SGPS
dataset we expect to catalogue many new shells.  How are these shells
distributed in the Galaxy?  Do they lie along spiral arms like GSH
277+0+36 and GSH 280+0+59, or is their distribution seemingly random
as in the Large and Small Magellanic Clouds (Kim et al.\ 1998;
Staveley-Smith et al.\ 1997)?  How large an impact do shells have on
the global ISM?  Will the Milky Way resemble the LMC and SMC which are
dominated by shells?

With respect to the small scale structure, we ask: How are the HISA
features distributed?  Do they lie predominantly along spiral shocks
or are they ubiquitous?  How do they relate to the molecular clouds?
What fraction of the ISM is in the CNM?  Are there larger, warmer
clouds that we are missing?  How small are the numerous clouds that
remain unresolved by the SGPS?

\section{Conclusions}
New large-scale, high-resolution \HI\ surveys like the SGPS and CGPS
finally make it possible to simultaneously study the small and large
scale structure of the ISM.  We hope that the SGPS dataset will allow
us to understand how the large and small scale structures relate to
each other throughout the inner Galaxy.  The SGPS should be able to
catalogue the majority of HI shells in the inner Galaxy and determine
exactly how they shape the ISM.  It is likely that we are currently
seeing only the most dramatic cold clouds with HISA, however further
studies may lead to a complete understanding the distribution of the
CNM.


%
%





\section*{Acknowledgements}
JMD and NMM-G acknowledge support of NSF grant AST-9732695 to the
University of Minnesota.  NMM-G is supported by NASA Graduate Student
Researchers Program (GSRP) Fellowship NGT 5-50250.  BMG acknowledges
the support of NASA through Hubble fellowship grant HST-HF-01107.01-A
awarded by STScI, which is operated by AURA Inc. for NASA under
contract NAS 5-26555.


\section*{References}
\reference Caswell, J. L. \& Haynes, R. F., 1987, A\&A, 171, 261

\reference Dickey, J.\ M., McClure-Griffiths, N., Gaensler, B., Green,
A., Haynes, R., Wieringa, M., 1999, in New Perspectives on the
Interstellar Medium, ASP Conf. Ser. 168, eds. A. R. Taylor,
T. L. Landecker \& G.  Joncas, (San Francisco: ASP), 27

\reference Fich, M., Blitz, L., \& Stark, A. A., 1989, ApJ, 342, 272

\reference Gibson, S. J., Taylor, A. R., Higgs, L. A., \& Dewdney,
P. E., 2000, ApJ, 540, 851

\reference Kim, S., Dopita, M. A., Staveley-Smith, L., Bessell, M. S.,
1998, AJ, 118, 2797

\reference McClure-Griffiths, N. M., Green, A. J., Dickey, J. M.,
Gaensler, B. M., Haynes, R. F., Wieringa, M. H., 2000a, ApJ, submitted

\reference McClure-Griffiths, N. M., Dickey, J. M., Gaensler, B. M.,
Green, A. J., Haynes, R. F., Wieringa, M. H., 2000b, AJ, 119, 2828

\reference Staveley-Smith, L., Sault, R. J., Hatzidimitriou, D.,
Kesteven, M. J., McConnell, D., 1997, MNRAS, 289, 225

\reference Taylor, A.\ R., 1999, in New Perspectives on the
Interstellar Medium, ASP Conf.\ Ser.\ 168, eds.\ A.\ R.\ Taylor, T.\
L.\ Landecker \& G.\ Joncas, (San Francisco: ASP), 3





\newpage
\begin{figure}[htb] 
\begin{center}
\caption{Grey-scale image of an \HI\ channel map at $v=2.12$ \kms\
  showing an apparent \HI\ shell approximately $2\fdg5$ in diameter in the
  local ISM at $l=330\fdg5$, $b=2\fdg2$.  The grey-scale is logarithmic to
  emphasize the shell walls.  The brightness temperature scale in Kelvins is
  displayed in the wedge on the right.}
\label{fig:local}            
\end{center}
\end{figure}
   
\begin{figure}[htb] 
\begin{center}
\caption{Grey-scale image of an \HI\ channel map at $v=-108$ \kms\
  showing an apparent \HI\ shell in the ISM. A small shell, of diameter
  $\sim 0\fdg4$ is located at $l=329\fdg3$, $b=0\fdg4$. This shell is
  identified by the bright ring of emission surrounding an \HI\ void.  The
  grey-scale is linear as shown in the wedge on the right.}
\label{fig:term}            
\end{center}
\end{figure}
 
\begin{figure}[htb] 
\begin{center}
\caption{Grey-scale image showing three orthogonal slices through
  the \HI\ cube containing the supershell GSH 277+0+36.  The greyscale is
  linear and runs from 0 K (white) to 35 K (black).  The largest panel is a
  channel map at $v=+36$~\kms.  The panel to the right is the {\em b-v}
  slice at the position marked by the cross-hair in the {\em l-b} slice.
  Similarly, the bottom figure is the {\em l-v} slice at the same position.
  Several chimney-like extensions to high latitudes are visible in both the
  channel map and {\em b-v} slice (McClure-Griffiths et al. 2000b).}
\label{fig:gsh277}            
\end{center}
\end{figure}
\begin{figure}[htb] 
\begin{center}
\caption{A similar image to Figure~\ref{fig:gsh277}, showing
  orthogonal slices through the center of the supershell GSH 280+0+59.}
\label{fig:gsh280}            
\end{center}
\end{figure}

\end{document}